\documentclass[12pt]{article}
\title
{\Large \bf High Precision Mass Measurements \\
in $\Psi$ and $\Upsilon$ Families Revisited}

\author
{A.S.Artamonov, S.E.Baru, A.E.Blinov, V.E.Blinov, \\
A.E.Bondar, A.D.Bukin, A.G.Chilingarov, N.F.Denisov, \\
S.I.Eidelman, Yu.I.Eidelman, V.R.Groshev, N.I.Inozemtsev, \\
G.Ya.Kezerashvili, V.A.Kiselev, S.G.Klimenko, G.M.Kolachev, \\
E.A.Kuper, L.M.Kurdadze, M.Yu.Lelchuk, S.I.Mishnev, \\
S.A.Nikitin, A.P.Onuchin, E.V.Pakhtusova, V.S.Panin, \\
V.V.Petrov, I.Ya.Protopopov, E.L.Saldin, A.G.Shamov, \\
Yu.M.Shatunov, B.A.Shwartz, V.A.Sidorov, Yu.I.Skovpen, \\
A.N.Skrinsky, V.A.Tayursky, V.I.Telnov, A.B.Temnykh, \\
Yu.A.Tikhonov, G.M.Tumaikin, A.E.Undrus, A.I.Vorobiov, \\
M.V.Yurkov, V.N.Zhilich, A.A.Zholentz\\
\\
\it Budker Institute of Nuclear Physics, 630090, Novosibirsk, Russia}

\begin{document}
\maketitle

\begin{abstract}
   High precision mass measurements in $\Psi$ and $\Upsilon$ families 
performed in 1980-1984 at the VEPP-4 collider with OLYA and MD-1 detectors 
are revisited. The corrections for the 
new value of the electron mass are presented. The effect of 
the updated radiative corrections has been calculated for the  $J/\Psi(1S)$ 
and $\Psi(2S)$ mass measurements.

\end{abstract}
\vspace*{7mm}
\rm

\par
   Development of the resonant depolarization method (RDM) suggested
in Novosibirsk \cite{bukin,derb} opened unique opportunities in the high
precision determination of the elementary particle masses.
Pioneer experiments in Novosibirsk (see \cite{ss} and references 
therein) were followed by those at Cornell \cite{cornell},
DESY \cite{desy} and CERN \cite{cern}. 
In this paper we reconsider our measurements performed at the 
e$^+$e$^-$ collider VEPP-4 in Novosibirsk in the $\Psi$ \cite{olya1,olya2} and 
$\Upsilon$ meson families \cite{md1,md2,md3,md4,md5} with the goal to take 
into account the change of the electron mass value \cite{emass73,emass87} 
as well as the updated radiative corrections \cite{kf} in case of $J/\Psi(1S)$ and $\Psi(2S)$.
  
\par
   $J/\Psi(1S)$ and $\Psi(2S)$ mass measurements \cite{olya1,olya2} were 
performed in 1980 with the OLYA detector \cite{olyadec} while the MD-1 group 
\cite{mddec} carried out three independent measurements of the 
$\Upsilon(1S)$ mass in 1982 \cite{md1}, in 1983 \cite{md2} and 
in 1984 \cite{md3,md4} as well as determined the masses of 
$\Upsilon(2S)$ and $\Upsilon(3S)$ in 1983 \cite{md2,md5}. The masses of the 
$\Psi$ and $\Upsilon$ mesons were obtained 
from a fit of the energy dependence of $\sigma(e^+e^- \to hadrons)$ 
and relating the value of the resonance mass to the beam energy.
The absolute calibration of the beam energy was performed using the RDM.

The resonant depolarization method is based upon the fact that in a storage 
ring with a planar orbit the spin precession frequency $\Omega_s$ depends 
on the beam energy $E$ as
\begin{equation}
\Omega_s=\omega(1+\frac{\mu'}{\mu_0}\gamma),
\end{equation}
where $\omega$ is the beam revolution frequency,
      $\mu'/\mu_0$ is the ratio of the anomalous and normal
      parts of the electron magnetic moment,
      $\gamma=E/m c^2$ is the Lorentz factor of electrons.
   The frequency $\Omega_s$ is measured at the polarized electron beam 
using a depolarizer with the frequency $\Omega_d$ adjusted as $\Omega_d = \Omega_s + n\omega$, where n is an arbitrary integer number.

\par
A typical accuracy of the method is about $10^{-5}$.
However, the measured quantity is a $\gamma$ factor of electrons rather 
than their energy. 
Thus, the beam energy and the resonance mass determined by the RDM 
depend on the electron mass assumed. 
In 1986 when the results of the $\Upsilon(1S)$ mass measurement were 
published \cite{md3}, its accuracy was about five times worse than the 
claimed accuracy of the electron mass in the MeV scale (2.8 {\it ppm}) 
\cite{emass73}.
However, in ``The 1986 adjustment of the fundamental physical constants'' 
\cite{emass87} the value of the electron mass was decreased by 8.5 {\it ppm}
while its error was reduced to 0.3 {\it ppm}.

\par
The decrease of the electron mass \cite{emass87} was caused mainly by 
the 7.8 {\it ppm} (about three ``old'' standard deviations) increase of 
the $e/h$ ratio. 
Taking into account that two other fundamental constants which 
depend on $e$, $h$ and $m_e$, i.e. the fine-structure constant $\alpha$ and 
Rydberg constant $R_{\infty}$, remained almost unchanged, the increase of 
$e/h$ propagates to the abovementioned 8.5 {\it ppm} decrease of $m_e$ 
in the MeV scale.
Since resonance masses determined from RDM are based upon the value of the 
electron mass and are quoted in MeV, they should be also decreased by 
8.5 {\it ppm}. The corresponding corrections to the values of the 
$\Upsilon(1S), \Upsilon(2S)$ and $\Upsilon(3S)$ meson masses measured 
by MD-1  were already reported at the Chicago Conference \cite{conf}.

\par
An additional 
correction should be applied to the values of the $J/\Psi(1S)$ and 
$\Psi(2S)$ mass obtained in  \cite{olya1,olya2}. 
Similarly to most early measurements, a fit of $\sigma(e^+e^- \to hadrons)$ 
in these papers 
included
the radiative corrections 
calculated according to the classic 
work of Jackson and Scharre \cite{js}. 
Later, in Ref.  \cite{kf} it was shown that
the approach of Ref. \cite{js} is not quite accurate  and,
in particular, violates the Bloch-Nordsieck theorem.
Correspondingly, the analysis of the $\Upsilon$ resonances 
was performed \cite{md3,md4,md5}
using the improved radiative corrections suggested in \cite{kf}.
In Ref. \cite{md3} it was shown that the corresponding shift of the
mass was about 0.1 MeV.
Somewhat later the paper  \cite{ale} was published 
entirely dedicated to the correction of the old measurements 
of $\Psi$ and $\Upsilon$ parameters using the updated radiative 
corrections.
However, the $J/\Psi(1S)$ and $\Psi(2S)$ masses were neither refit 
by the authors of Ref.\cite{olya1,olya2} nor quoted in Ref.\cite{ale}.

\par
The details of $J/\Psi(1S)$ and $\Psi(2S)$ mass measurements 
\cite{olya1,olya2} are not available now.
Therefore, the $J/\Psi(1S)$ and $\Psi(2S)$ mass corrections were estimated 
by us as in Ref.\cite{ale} from the difference of the fits with 
the radiative corrections from Ref.\cite{js} and Ref.\cite{kf}.
Similarly to Ref.\cite{js}, only the electron loop was taken into account 
in the photon vacuum polarization term in Ref.\cite{olya1,olya2}.
The resulting mass correction for radiative effects equals -(0.023$\pm$0.003)$\sigma_w$,
where $\sigma_w$ is the rms spread of the $e^+e^-$ center of mass energy and 
the error accounts for dependence of the correction on the luminosity 
distribution around the resonance. 
The correction is somewhat lower than that which can be obtained from Fig.6 
of Ref.\cite{ale}. At $\sigma_w$ = 0.7(1.0) MeV in $J/\Psi(1S)$ and 
$\Psi(2S)$ runs it equals -0.016 MeV and -0.023 MeV respectively.

\par
 Table 1 presents a list of the resonance masses 
measured at the VEPP-4 collider with the corresponding corrections,
where $\Delta{M}(m_{e})$ and $\Delta{M}(rad.)$ stand for the correction
for the electron mass and radiative effects respectively.

\par
Let us briefly discuss how the change of the resonance masses above
can affect other measurements. The new value of the $\psi(2S)$ mass should be
taken into account during the interpretation of the Fermilab studies of the
charmonium family in $p\bar{p}$ annihilation \cite{armstrong} which used 
the value of the $\psi(2S)$ mass from \cite{olya1,olya2} as a 
basic calibration in their determination of the $J/\psi(1S)$ mass.
It is obvious that the obtained values of the $m_e$ correction for
$\Upsilon(1S)$ and $\Upsilon(2S)$ can also be applied to the Cornell 
\cite{cornell} and DESY \cite{desy} measurements respectively.  
Since in these experiments the radiative corrections were calculated
according to Ref.\cite{js}, their results should be also corrected for the
radiative effects.
We remind that our value of the $\Upsilon(1S)$ mass differs by more than
3.5 standard deviations from that at Cornell while for
$\Upsilon(2S)$ it is consistent with the one in DESY.
Our measurement of the $\Upsilon(3S)$ mass has not been repeated by any
other group.




\begin{table}
\caption{Revision of mass measurements in 
$\Psi$ and $\Upsilon$ families}
\label{tab:1}
\begin{center}
\begin{tabular}{|c|c|c|c|c|}
\hline
{\bf Particle} & {\bf Previous mass,} & {\bf  $\Delta{M}(m_e)$,} & 
{\bf  $\Delta{M}(rad.)$,} & {\bf Updated mass,} \\
& {\bf MeV}  & {\bf MeV}  & {\bf MeV} & {\bf MeV}  \\
\hline
$J/\Psi(1S)$  \cite{olya1,olya2}  & 3096.93$\pm$0.09 & -0.026 & -0.016 &
 3096.89$\pm$0.09 \\
\hline
$\Psi(2S)$  \cite{olya1,olya2}   &    3686.00$\pm$0.10 & -0.031 & -0.023 &
 3685.95$\pm$0.10 \\
\hline
$\Upsilon(1S)$ \cite{md4} & 9460.59$\pm$0.09$\pm$0.05 &  -0.080  & - &
 9460.51$\pm$0.09$\pm$0.05 \\
\hline
$\Upsilon(2S)$ \cite{md2,md5} & 10023.6$\pm$0.5 &  -0.085  & - & 
10023.5$\pm$0.5 \\
\hline
$\Upsilon(3S)$ \cite{md2,md5} & 10355.3$\pm$0.5 &  -0.088  & - &
 10355.2$\pm$0.5 \\
\hline
\end{tabular}
\end{center}
\end{table}

\par

\end{document}